\title{Dwell time of a Brownian interacting molecule in a cellular microdomain}
\author{Adi Taflia\thanks{Department of Applied Mathematics, Tel-Aviv University, Ramat-Aviv
69978 Tel-Aviv, Israel.}\,  and David Holcman \thanks{Department of
Mathematics, Weizmann Institute of Science, Rehovot 76100, Israel. }
}
\documentclass[12pt,date,simplespacing]{article}
\usepackage{epsfig, graphicx}
\usepackage{setspace}
  \usepackage{a4wide}
\usepackage{float}
   \usepackage{amsmath}
\usepackage{latexsym}
\usepackage{epsfig, graphicx} 

\newcommand{\p}{\partial}
\newcommand{\ds}{\displaystyle}
\newcommand{\beq}{\begin{eqnarray}}
\newcommand{\beqq}{\begin{eqnarray*}}
\newcommand{\eeq}{\end{eqnarray}}
\newcommand{\eeqq}{\end{eqnarray*}}
\newcommand{\eps}{\varepsilon}
\newcommand{\x}{\mbox{\boldmath$x$}}
\newcommand{\y}{\mbox{\boldmath$y$}}

\newcommand{\J}{\mbox{\boldmath$J$}}
\newcommand{\n}{\mbox{\boldmath$n$}}
\newcommand{\mean}[1]{\langle #1\rangle}

\newtheorem{prop}{Proposition}

\def\ds#1{\displaystyle{#1}}

\def\QED{\quad\hbox{\hskip 4pt\vrule width 5pt height 6pt depth 1.5pt}}

\begin{document}
\numberwithin{equation}{section}
\maketitle
\begin{abstract}
The time spent by an interacting Brownian molecule inside a bounded
microdomain has many applications in cellular biology, because the
number of bounds is a quantitative signal, which can initiate a
cascade of chemical reactions and thus has physiological
consequences. In the present article, we propose to estimate the
mean time spent by a Brownian molecule inside a microdomain $\Omega$
which contains small holes on the boundary and agonist molecules
located inside. We found that the mean time depends on several
parameters such as the backward binding rate (with the agonist
molecules), the mean escape time from the microdomain and the mean
time a molecule reaches the binding sites (forward binding rate). In
addition, we estimate the mean and the variance of the number of
bounds made by a molecule before it exits $\Omega$. These estimates
rely on a boundary layer analysis of a conditional mean first
passage time, solution of a singular partial differential equation.
In particular, we apply the present results to obtain an estimate of
the mean time spent (Dwell time) by a Brownian receptor inside a
synaptic domain, when it moves freely by lateral diffusion on the
surface of a neuron and interacts locally with scaffolding
molecules.
\end{abstract}

\section{Introduction}
Biochemical reactions in cellular microdomains involve, in general,
a small number of molecules that can bind to agonist molecules,
confined in some subregions. A microdomain $\Omega$ is defined as a
bounded domain where a large fraction of the boundary $\p\Omega_r$
is reflective and a small part $\p\Omega_a$ of it is absorptive,
which allows molecules to enter or/and exit. Under normal
physiological conditions, a molecule can be trapped inside a
microdomain for a period long enough, compared to other time scales
such as free diffusion or binding time. Because a molecule can be
trapped for a long time, many chemical bounds can be made, before it
exits. From a physiological stand point, an interesting property of
such  microdomain is that depending on the number of chemical bounds
made by a molecule, a cascade of chemical reactions can be
initiated, which ultimately affects some physiological properties.
Our interest here is to estimate the number of chemical bounds made
by a Brownian interacting molecule inside $\Omega$. This number
depends on the geometry of the microdomain and the distribution of
agonist molecules. We also provide an estimation of the mean time
spent (Dwell time) by a molecule inside the microdomain, denoted by
$E(\tau_D)$. For that purpose, we derive an asymptotic formula for
the Dwell time, the mean and the variance of the number of bounds
made before the molecule exits, as the ratio
$\frac{|\p\Omega_a|}{|\p\Omega_r|}$ tends to zero. Finally, we
derive a formula for the Dwell time of a molecule inside $\Omega$,
when a steady state flux of molecules is maintained fixed at the
absorbing boundary. The role of the flux is to fix the number of
free binding sites. Using the present method, we also obtain an
estimate of the forward binding rate constant. Historically, the
theory of chemical reactions at a molecular level limited by
diffusion has been developed by many authors, to quote but a few
\cite{Wilemski,Perico,Agmon,Szabo1,Szabo2}: using the classical
theory of diffusion and interactions with binding sites, various
rate constants were computed. Recently, using averaged equations,
chemical reactions in microdomains have been described in
\cite{BereSmalhole,Berepartialflux}. Chemical reactions in closed
microdomains were studied in \cite{JChemPhys}, where we obtain, in
particular, some estimates on the mean of the variance of the
number of bound molecules in a steady state regime. In \cite{HS},
an asymptotic estimate of the mean time it takes for a Brownian
molecule to escape an empty domain through small openings located
on the boundary, was obtained using a new type of singular
perturbation problem. More specifically, if D denotes the
diffusion constant, $|\Omega|$ the volume of the domain $\Omega$
and $\varepsilon=\frac{|\p \Omega_a|}{|\p \Omega|}<<1$ is the
ratio of the absorbing to the total boundary, then for
$\varepsilon$ small, the leading order term of the mean escape
time $\tau(\x)$ (for a molecule starting at position $\x$, far
from the entrance) is given by
\beq\label{HolcSchuss}
\tau(\x) =
\frac{|\Omega|}{\pi D}\log(\frac1{\varepsilon})+O(1).
\eeq
In the first approximation, the mean time $\tau(\x)$ does not depend
on the initial position $\x$ and will be denoted by $\tau$. In this
article, we obtain an explicit asymptotic estimation of the Dwell
time $E(\tau_D)$ as a function of the characteristic sizes of the
domain $\Omega$, the size of the small openings $\p \Omega_a$, the
number of the binding molecules. More specifically, $E(\tau_D)$ is
given by expression (\ref{eq:formula_mean_mean}), which depends on
the mean time $\mean{\tau}$ to exit when no binding occurs, the mean
time $\mean{T}$ to enter into the binding site area, $m_{\delta}$
the mean probability to bind before exit and the backward binding
rate $k_{-1}$. We get
\begin{equation}\label{eq:formula_mean_meanintro}
E(\tau_D) = \mean{\tau}+\frac{1-m_{\delta}}{m_{\delta}}\left(
\mean{T}+\frac{1}{k_{-1}}\right).
\end{equation}
It is well known from the theory of chemical reactions that the
backward binding rate $k_{-1}$ depends only on the local
interactions between two interacting molecules. If $\Delta E$
denotes the activation barrier, $kT_e$ is the energy due to the
temperature, the Arrhenius law states that: \beq k_{-1} = C
e^{-\frac{\Delta E}{kT_e}},
 \eeq
where C is a constant that depends on the temperature $T_e$, the
electrostatic potential barrier $\Delta E$ generated by the
binding molecule and the friction coefficient \cite{Schuss}. In
the first part of the paper, we derive equation
(\ref{eq:formula_mean_meanintro}) by counting the number of bounds
between the Brownian molecule and the agonist molecules, before
the Brownian molecule exits the domain. In the second part, we
derive some asymptotic estimates of the quantities $\mean{T}$,
$\mean{\tau}$ and $m_{\delta}$ as a function of the geometry, when
the radius $\delta$ of the binding site tends to zero. Although
the present computations are carried out in two dimensions, they
can be extended to dimension 3 by using the techniques developed
in \cite{SSHE}. Finally in the last part, we apply the present
computations to study chemical reactions occurring in synaptic
microdomains: the chemical reactions are the binding of receptors
with the scaffolding molecules. It is indeed of great interest to
analyze the mechanism that regulates the number and the type of
receptors at a synapse, because receptors control the synaptic
weight. Any fluctuations of the number results in a variation of
the synaptic weight and affects the reliability of the synaptic
transmission. Moreover, certain experimental protocols have lead
to a Long Term Potentiation of a synapse, a mechanism which is
associated with a change of the number and the type of certain
receptors \cite{Bredt,Nicoll}. The regulation of synaptic
plasticity is a fundamental process underlying learning and memory
\cite{Bredt,Nicoll} and recently, single molecule tracking has
revealed that the number of postsynaptic receptors, which
participate to the synaptic transmission, is not fixed but it
changes due to constant traffick of receptors on the surface of
neurons. Receptors move in and out from synaptic regions
\cite{Choquet} \cite{Triller} and following these observations,
many questions have been raised: in particular, what determines
the time spent by a receptor inside a synapse? How receptors can
be stabilized inside a synapse? How long they stay inside synaptic
microdomains? Such questions are partially answered in the present
paper. In particular, our computation of the Dwell time of a
receptor inside a specific microdomain, called the Postsynaptic
density (PSD) takes into account the interaction with the
scaffolding molecules, which was ignored in \cite{HS}.

\subsection{Molecular dynamics in a microdomain}
{The dynamics of a molecule or a protein moving on the surface of
a cell is usually described in the Smoluchowsky limit (large
friction) of the Langevin equation \cite{Schuss}: for a molecule
of mass $m$, described by its position $X$ at time $t$, with a
friction coefficient $\gamma$, moving inside a potential well $V$,
the Smoluchowsky limit of the Langevin equation is \beq\label{SL}
\gamma \dot{X}+\nabla V(X)=\sqrt{2\gamma \varepsilon_e}\dot{w},
 \eeq
where $\varepsilon_e=\frac{kT_e}{m}$ and $w$ is a Gaussian random
variable with variance 1 and mean 0. In a microdomain $\Omega$,
where a large fraction of the boundary is reflective $\p \Omega_r$
and a small part of it is absorptive $\p \Omega_a$, the probability
density function (pdf) $p$ to find $X$ at time $t$ in the surface
element $\x+d\x$ satisfies the Fokker-Planck Equation (FPE)
\beq\label{FPE} \frac{\p p(\x,t)}{ \p t} &=& D \Delta p(\x,t) -
\nabla\cdot (\nabla V(\x)p(\x,t)) \hbox{ for } \x \in \Omega\\
& & \nonumber\\
\J(\x,t)\cdot\n&=&0\hbox{ for } \x \in \p \Omega_r \\
& & \nonumber\\
p(\x,t)&=&0 \hbox{ for } \x \in \p \Omega_a
 \eeq
where $D=\gamma \varepsilon_e$ is the diffusion constant, $\n$ is
the external normal at the boundary, the flux $\J$ is given by
\beq \J(\x,t)=-D\nabla p(\x,t) +\nabla V(\x)p(\x,t).
 \eeq
We denote by $t^{x}$ the first time the molecule arrives at the
absorbing boundary  $\p \Omega_a$, when it started at position $\x$.
It is well known \cite{Schuss} that the mean first passage is the
expectation of the time $t^{x}$ and is given by
\beqq
E^{\x}(t^{x})&=& \int_{0}^{\infty} t\frac{d}{dt}
 Pr\{ t^x<t \}dt= \int_0^{\infty} Pr\{t^x>t\}dt\\
           &=& \int_0^{\infty} \int_{\Omega} p(\y,t|\x)d\y dt,
\eeqq
where $p(\y,t|\x)$ is the pdf of the process $X$, conditioned on the
initial position $\x$, that is, as $t$ goes to zero, \beq
p(\y,t|\x)\rightarrow \delta(\x-\y),
 \eeq
where $\delta$ is the Delta-Dirac function. In equation (\ref{FPE}),
$V$ represents the potential wells generated by the binding
molecules inside the domain $\Omega$. It is, in fact, the sum of the
local potential wells, supported inside a ball of finite radius
generated by the binding molecules. Usually the binding molecules
are scattered inside the domain $\Omega$, but in the present model
we replace the scattered distribution of binding molecules by a
simplified distribution, where we imagine that all the binding
molecules are located inside a compartment $D(\delta)$ in $\Omega$.
In the present description, the potential $V$ becomes an effective
potential defined in $D(\delta)$, whose characteristics should be
such that the classical chemical reaction theory is recovered. More
specifically,  we can define the microdomain $\Omega$ containing the
binding domain $D(\delta)$, which replaces the domain with many
scattered binding sites: this simplified domain made of two
compartments is called the homogenized microdomain and is described
as (see figure 1)
\begin{enumerate}
\item An internal compartment, which is described as a disk
$D(\delta)$ of radius $\delta$. This disk represents the region
where the binding sites are located. Instead of using the dynamics
associated with equation (\ref{SL}), we describe the entrance and
the exit of a molecule inside $D(\delta)$ using a Poissonnian
description, where the mean can be related to the properties of
the potential well. For that purpose, we recall that a chemical
reaction with a binding molecule is described as the arrival of a
Brownian molecule inside the disk $D(\delta)$. The release process
is modeled as the escape of the molecule from the potential well
$V$ and is described by the chemical reaction \beq \label{chem}
R+S
\begin {array}{c}
{k_1}\\
\rightleftharpoons \\
{k_{-1}} \end{array} RS \eeq where  $\frac{1}{k_{-1}}$ is the mean
binding time and depends only on the local potential well, generated
by the binding molecules \cite{Tier84} \cite{Schuss}.
\item An external compartment separated from the rest of the
biological environment by the boundary $\p \Omega_r$, containing
small openings $\p \Omega_a$. Molecules can enter or exit through
the openings and thus can be exchanged with the rest of the
cellular medium (see figure 1). The dynamics of a molecule in that
compartment is described as pure Brownian until it escapes.
\end{enumerate}
\begin{figure}[htbp]\label{fig1}
\centerline{\includegraphics[width=2.0in,height=1.9in]{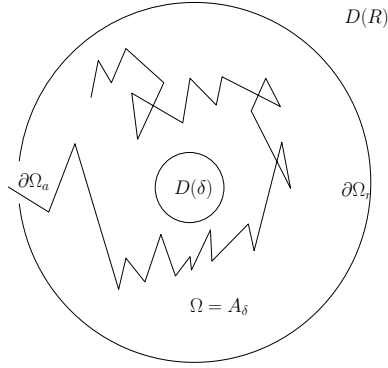}}
 \caption{Model of a cellular microdomain in two
dimensions. The domain is a disk $D(R)$ of radius $R$, made of two
compartments: an inner disk $D(\delta)$ of radius $\delta$ and the
annulus $A_\delta=D(R)-D(\delta)$. A molecule moves by Brownian
motion inside $A_\delta$ until it hits $D(\delta)$ or the
absorbing boundary $\p \Omega_a$. When the Brownian molecule
enters into $D(\delta)$, which represents the domain of chemical
reactions, while it reacts with an effective binding molecule
during a mean time (the inverse of the backward binding rate) its
movement in $D(\delta)$ is frozen. The molecule is then released
uniformly inside the annulus $A_\delta=D(R)-D(\delta)$. This
scenario repeats until the molecule hits the absorbing boundary
$\p \Omega_a$, where the molecule is finally removed.}
\end{figure}

\subsection{Time spent by a molecule inside a disk containing a small opening}
We estimate asymptotically the Dwell time $E(\tau_D)$ of a molecule
inside a domain $\Omega$, when the domain $\Omega$ is a disk $D(R)$
of radius $R$. The case of a general domain is left open. By
definition, $E(\tau_D)$ is the mean time to exit, averaged over an
initial uniform distribution. We make the following assumptions: the
ratio $\varepsilon$ of the absorbing to the reflective boundary of
$D(R)$ is small $<<1$. The absorbing (resp. reflecting) part is
denoted $\p\Omega_a$ (resp. $\p
\Omega_r$). The computation of the Dwell time will be made using
the homogenization version of the domain, described in the
previous paragraph: the binding molecules are located in the small
concentric disk D($\delta$), where $\delta<<R$. Each time a
molecule enters into D($\delta)$, it can be bound with a binding
molecule and will be released in the annulus
$A_{\delta}(R)=D(R)-D(\delta)$. Since R is a fixed parameter, we
will denote the annulus by $A_{\delta}$.
 We assume that the boundary $\p D(\delta)$ is always absorbing.
\section{The Dwell Time $E(\tau_D)$}
 The goal of this section is to derive a general expression of the Dwell time
$E(\tau_D)$. The movement of a molecule is described here as
Brownian (see \cite{Choquet1,Choquet} for the case of a synapse)
with a diffusion constant $D$ and $X^{\x}(t)$ is the position of a
molecule at time $t$ starting at position $\x$. $E(\tau_D)$ is the
mean time a molecule remains inside the domain $\Omega$ before it
exits, averaged over a uniform initial distribution inside the
annulus $A_{\delta}=D(R)-D(\delta)$. Inside  $A_{\delta}$, a
 molecule diffuses freely until it hits $\p D(\delta)$. Inside $D(\delta)$, a bound
 molecule is kept fixed for a mean time $\frac1{k_{-1}}$, before being released inside
 $A_{\delta}$ at a random position, distributed uniformly.
To estimate the Dwell time $E(\tau_D)$, we first consider the
Dwell time $E(\tau^{\x})$, conditioned on the initial position
located inside $A_{\delta}$. We do not consider here the entrance
of the Brownian receptor inside the domain, which would require to
use a Langevin formulation of the dynamics \cite{Schuss}. We
derive a formula for $E(\tau^{\x})$ by counting the number of
times a molecule enters into $D(\delta)$ before it exits: either a
molecule exits with no bindings or many bindings occur, before the
receptor leaves $D(R)$. Let $T^A_{\x}$ denote the first passage
time the trajectory $X^{\x}(t)$ hits the absorbing boundary
$\p\Omega_a$. Similarly we define by $T^S_{\x}$ the first passage
time of the trajectory $X^{\x}(t)$ to the boundary of $D(\delta)$.
  We define now the probability $p_{\delta}(\x)$ that a
trajectory $X^{\x}(t)$ leaves the domain $A_{\delta}$ before
 any bounds occur, explicitly
\beq p_{\delta}(\x)=\Pr\{T^A_{\x}<T^{S}_{\x}\}. \eeq
\subsection{The general formula for the mean Dwell time $E(\tau_D)$}
To estimate $E(\tau_D)$, we consider the different random events
related to the number of times
 the Brownian molecule enters into the region $A_{\delta}$, before it exits.  $F^{\x}_0$ is the event that a
receptor exits the microdomain $A_{\delta}$ when no bindings
occur, starting at position $\x$. $F^{\x}_1$ is the event that a
receptor starting at $\x$, enters the domain $D(\delta)$ only once
and then exits without entering again in $D(\delta)$. Similarly,
we define the event $F^{\x}_n$ that a receptor starting at
position $\x$ exits after exactly $n$ bounds. The probability that
the molecule has left the domain before time $t$ is
$Pr\{\tau^{\x}<t\}$. The mean time is given by \beq
E(\tau^{\x}) &=&\int_{0}^{\infty} t\frac{d}{dt} Pr\{\tau^{\x}<t\}dt\\
          &=& \sum_n\int_{0}^{\infty} t\frac{d}{dt}
          Pr\{\tau^{\x}<t,F^{\x}_n\}dt.
\eeq We use Bayes formula to get \beq \label{mfpt}
E(\tau^{\x})=E(\tau^{\x}|F^{\x}_0)\Pr(F^{\x}_0)+E(\tau^x|F^{\x}_1)\Pr(F^{\x}_1)+E(\tau^{\x}|F^{\x}_2)\Pr(F^{\x}_2)+....
\eeq where \beq E(\tau^{\x}|F^{\x}_n) =\int_{0}^{\infty}
t\frac{d}{dt} Pr\{\tau^{\x}<t|F^{\x}_n\}dt. \eeq
$Pr\{\tau^{\x}<t|F^{\x}_n\}$ is the probability that $\tau^{\x}<t$
conditioned to the event that exactly $n$ bounds are made before the
molecule exits the domain $D(R)$. When a molecule detaches from the
domain $D(\delta)$, in this model, it is released inside the domain
$A_{\delta}$ according to a density distribution $\rho(\x)$:
immediately after the mean duration
 $\frac1{k_{-1}}$, the molecule is released inside the volume element $\x+d\x$
 with a probability $\rho(\x)d\x$.
To compute the first term in equation (\ref{mfpt}), we notice that
\beq E(\tau^x|F^{\x}_0)\Pr(F^{\x}_0)
=E(T^A_{\x}|T^{A}_{\x}<T^S_{\x}) p_{\delta}(\x). \eeq The second
term $E(\tau^{\x}|F^{\x}_1)\Pr(F^{\x}_1)$ is the sum of two terms.
First the mean time it takes for a molecule to reach the boundary
$\p D(\delta)$ and stays there for a mean time $\frac1{k_{-1}}$ and
second the mean time to go from a point $\x_1$, where the molecule
starts afresh, to the absorbing boundary. If we note that the
probability to reach $\p D(\delta)$ before $\p\Omega_a$ starting
from $\x$ is $1-p_{\delta}(\x)$, then we have: \beqq & &
E(\tau^{\x}|F^{\x}_1)\Pr(F^{\x}_1) =(1-p_{\delta}(\x))\times\\ & &
\left( (E\{T^S_{\x}|T^S_{\x}<T^A_{\x}\}+ \frac1{k_{-1}})
\int_{A_{\delta}} \rho(\x_1)p_{\delta}(\x_1)d\x_1 +
\int_{A_{\delta}} E\{T^A_{\x_1}|T^A_{\x_1}<T^S_{\x_1}\}
\rho(\x_1)p_{\delta}(\x_1)d\x_1 \right). \eeqq Similarly, the
general term where exactly $k$ bounds are formed is computed, with
the notation \beq \label{def-time}
t^A(\x)&=&E\{T^A_{\x}|T^A_{\x}<T^S_{\x}\}\\
& &  \nonumber \\
 t^S(\x)&=& E\{T^S_{\x}|T^S_{\x}<T^A_{\x}\}. \eeq
 We get,
\beqq
E(\tau^{\x}|F^{\x}_n)\Pr(F^{\x}_n)
&=&\int_{A_{\delta}}...\int_{A_\delta}
\left( \sum_{i=0}^{n-1}t^{S}(\x_i)+\frac{n}{k_{-1}}+t^A(\x_n) \right)\\
& & \prod_{i=1}^{n-1}(1-p_{\delta}(\x_i))\rho(\x_i)(1-p_{\delta}
(\x))p_{\delta} (\x_n)\rho(\x_n)d\x_1\ldots
d\x_n.\\
 &=& (1-p_{\delta}(\x)) \Big{\{}t^S(\x)(1-m_{\delta})^{n-1}m_\delta \\
 &+&(n-1) (1-m_{\delta})^{n-2}m_\delta\int_{A_{\delta}}t^S(\y)(\rho(\y)-p_{\delta}(\y)\rho(\y))d\y \\
&+&\frac{n}{k_{-1}}(1-m_{\delta})^{n-1}m_\delta+(1-m_{\delta})^{n-1}
\int_{A_{\delta}} t^A(\y)p_{\delta}(\y)\rho(\y)d\y \Big{\}} \eeqq
where
\beqq m_{\delta}= \int_{A_{\delta}}p_{\delta}(\y)\rho(\y)d\y .\eeqq
Formula (\ref{mfpt}) is in fact a geometric sum and using the
generic expression of the series, we get \beq\label{eq:mean_general}
E(\tau^{\x}) &=& t^A(\x)
p_{\delta}(\x)+\frac{1-p_{\delta}(\x)}{m_\delta}\Big{\{}t^S(\x)+\int_{A_{\delta}}t^A(\y)
p_{\delta}(\y)\rho(\y)d\y\\
&+& \frac1{m_\delta} (\int_{A_{\delta}}t^S(y)
(\rho(\y)-p_{\delta}(\y)\rho(\y))d\y +\frac{1}{k_{-1}})\Big{\}}
\nonumber . \eeq When $\rho$ is uniformly distributed, formula
(\ref{eq:mean_general}) can be simplified. Using that \beqq
m_{\delta}=\frac{1}{|A_{\delta}|}\int_{A_{\delta}}p_{\delta}(\y)d\y,
\quad\quad \eeqq where $|A_\delta|$ is the volume of $A_\delta$ and
if we define the two mean times
\begin {equation}\label{eq:tau} \mean{\tau}=
\frac{\int_{A_{\delta}} t^A(\x) p_{\delta}(\x) d\x
}{\int_{A_{\delta}}p_{\delta}(\y)d\y}
\end{equation}
and
\begin{equation}\label{eq:T}
\mean{T}= \frac{\int_{A_{\delta}} t^S(\x)(1- p_{\delta}(\x))
d\x}{\int_{A_{\delta}}(1-p_{\delta}(\y))d\y },
\end{equation}
then the Dwell time $E(\tau_D)$ is given by the average of
$E(\tau^{\x})$ with respect to the uniform distribution and we get
\begin{equation}\label{eq:formula_mean_mean}
E(\tau_D) = \mean{\tau}+\frac{1-m_{\delta}}{m_{\delta}}\left(
\mean{T}+\frac{1}{k_{-1}}\right).
\end{equation}
\bigskip
In the next paragraphs we obtain an explicit asymptotic estimate
of for component of formula (\ref{eq:formula_mean_mean}).

{\noindent \bf Remark.} It seems artificial to free a bounded
molecule not immediately at the boundary of the domain $D_{\delta}$.
But in fact, the release process can be modeled by re-initiating the
molecule at the boundary of $D_{\delta}$, otherwise it would
immediately return to a bounded state. To avoid this unrealistic
behavior, we have used the distribution function $\rho(\y)$ to model
the releasing process. A more accurate scenario  would require to
use a Langevin description, which accounts for the acceleration of
the molecule: when a molecule is released with a random Gaussian
initial velocity, it can travel up to a certain distance, before its
velocity reaches the value of mean thermal velocity of free
molecules.
\subsection{The mean number of bounds before exit}
To estimate the mean number of bounds made by a molecule before it
exits, we use the analysis of the previous paragraph. The
probability that a molecule starting at $\x$ does not bind before
exit is exactly $\Pr(F^{\x}_0)$, while the probability that
exactly $k$ bounds are made is $\Pr(F^{\x}_k)$. We define the
average probability by \beqq \Pr(F_k)=\int_{A_{\delta}}
\Pr(F^{\x}_k) \rho(\x)d\x. \eeqq The probability $\Pr(F^{\x}_k)$
can be expressed in terms of the conditional probability $p$ as
follows \beqq \Pr(F^{\x}_1)= \int_{A_{\delta}} p_{\delta}(\y)
\rho(\y)d\y (1-p_{\delta}(\x)) \eeqq and the average probability
$\Pr(F_1)$ is given by \beqq \Pr(F_1)= \int_{A_{\delta}}
\rho(\x)\Pr(F^{\x}_1)d\x=\int_{A_{\delta}} \int_{A_{\delta}}
p_{\delta}(\y) \rho(\y)d\y \rho(\x)(1-p_{\delta}(\x)) d\x =
m_{\delta}(1-m_{\delta}). \eeqq More generally, \beqq \Pr(F_k)=
\int_{A_{\delta}}...\int_{A_{\delta}}p_{\delta}(\y_k)
\rho(\y_k)d\y_k (p_{\delta}(\y_{k-1}) \rho(\y_{k-1})dy_{k-1}) ...
(1-p_{\delta}(\y_1)) \rho(\y_1)d\y_1 (1-p_{\delta}(\x)) \eeqq and
after some integrations we get
 \beqq \Pr(F_k)=
m_{\delta}(1-m_{\delta})^k. \eeqq The mean $M_b$ and the variance
$V_b$ of the number of bounds before exit are given by the well
known formula of geometric probability, \beq \label{mbb}
 M_b&=& \sum k \Pr(F_k)= \sum km_{\delta}(1-m_{\delta})^k \nonumber\\&=&\frac{1-m_{\delta}}{m_{\delta}}
=\frac{\int_{A_{\delta}}(1-p_{\delta}(\y))d\y}{\int_{A_{\delta}}p_{\delta}(\y)d\y} \\
V_b&=& \sum k^2 \Pr(F_k)-(\sum k \Pr(F_k))^2
=\left(\frac{1-m_{\delta}}{m^2_{\delta}}\right). \eeq Later on we
will give
 an asymptotic expansion of $m_{\delta},V_b$ and $M_b$ as a
function of the parameter $\delta$.
\section{Estimation of the probability\\ $p_{\delta}(x)=Pr\{T^a_{\x}<T^{S}_{\x}\}.$}
To obtain an asymptotic expansion of the probability $p_{\delta}(x)$
that a molecule exits before it enters into the domain $D(\delta)$,
that is, before it binds to any binding sites, we use the notations
of the previous section. $T_x^a $ denotes the first time a molecule
starting from $x$ hits the absorbing boundary. We assume that when a
molecule exits the domain $\Omega$, it does not come back. $T_x^S$
is the first time a molecule hits the inner circle $\p D(\delta)$.
%
The probability
$q_{\delta}(\x)=1-p_{\delta}(\x)=\Pr\{T^{S}_{\x}<T^{a}_{\x}\}$ of
the event $T_{\x}^a>T_{\x}^S $ satisfies an elliptic partial
differential equation \cite{Karlin}, with mixed boundary conditions
given by  \beq \label{eq5}
 \Delta q_{\delta}&=&0\mbox{ on }A_{\delta} , \\
 \frac{\partial q_{\delta}}{\partial n}(\x)&=&0\mbox{ on }\partial \Omega _r , \nonumber \\
 q_{\delta}(\x)&=&0\mbox{ on }\p \Omega_a ,\nonumber\\
 q_{\delta}(\x)&=&1\mbox{ on }\p D(\delta), \nonumber
\eeq where $\p \Omega _a $ is the small opening located on the
external boundary of $\Omega$, $\partial \Omega _r $ is the
remaining part of the external boundary, which is reflecting. In
polar coordinates $(r,\theta)$, the portion of the boundary $\p
\Omega_a$ is parameterized by $|\theta-\pi|\leq\eps$. By using the
particular geometry of the annulus $A_{\delta}$, an explicit
 estimation of the solution can be obtained by using spectral methods, developed in the context
 of mixed boundary value problems \cite{Sneddon,Fabrikant1,Fabrikant2}. Using the method of
separation of variables, the solution $q_{\delta}$ of problem
(\ref{eq5}) has the general form \beq
&&q_{\delta}(r,\theta)=\frac{a_0}{2}+\sum_{n=1}^{\infty}
\Big[a_n\Big( \frac r R\Big)^n+b_n\Big(\frac \delta
r\Big)^n\Big]\cos(n\theta)+\alpha \log\Big(\frac r
\delta\Big).\label{eq:series} \eeq We wish now to estimate the
coefficients $a_n$ and $b_n$. We denote $\beta=\frac{\delta}{R}$.
From the boundary conditions on
 $\p D(\delta)$ and on $r=R$,  we get
\beq
 \frac{a_0}{2}+\sum_{n=1}^\infty
\Big[a_n\Big(\frac{\delta}{R}\Big)^n+b_n\Big]\cos(n\theta)&=1,& \label{bc1}\\
\sum_{n=1}^\infty
n\Big[\frac{a_n}{R}-\frac{b_n}{R}\beta^{n}\Big]\cos(n\theta)+
\frac{\alpha}{R}&=0&\mbox{ for $|\theta-\pi|>\eps$}\label{eq:ser1p}\\
1+\sum_{n=1}^\infty \Big[a_n+b_n\beta^{n}\Big]\cos(n\theta)-\alpha
\log(\beta)&=0&\mbox{ for $|\theta-\pi|\leq\eps$}\label{eq:ser2p}.
 \eeq
Using equation (\ref{bc1}), we get the following identities:
 \beqq
&&a_0=2\\
&& b_n=-a_n\beta^n \mbox{ for $n\geq 1$}. \eeqq Using the
identities above and (\ref{eq:ser1p}) and (\ref{eq:ser2p}), we
obtain the double series equations
 \beq
&\sum_{n=1}^\infty
n\Big[\frac{a_n}{R}+\frac{a_n}{R}\beta^{2n}\Big]\cos(n\theta)+
\frac{\alpha}{R}=0,&\mbox{ for $|\theta-\pi|>\eps$}\label{eq:ser1}\\
&1+\sum_{n=1}^\infty
\Big[a_n-a_n\beta^{2n}\Big]\cos(n\theta)-\alpha
\log(\beta)=0,&\mbox{ for $|\theta-\pi|\leq\eps$.}\label{eq:ser2}
 \eeq
Substituting $c_n=a_n(1+\beta^{2n})$ and $H_n=\frac{2\beta^{2n}}{1-\beta^{2n}}$
 equations (\ref{eq:ser1}),(\ref{eq:ser2}) have the following
 form
 \beq
&\frac{c_0}{2}+\sum_{n=1}^\infty \frac{c_n}{1+H_n}
\cos(n\theta) =0,&\theta\in [\pi,\pi-\eps]\label{eq:gen1}\\
&&\nonumber\\
 &\alpha+\sum_{n=1}^\infty n c_n
\cos(n\theta)=0,&\theta\in[0,\pi-\eps],\label{eq:gen2} \eeq where
\beq\label{eq:c0} \frac{c_0}{2}=1-\alpha\log(\beta). \eeq The
asymptotic solution of  equations (\ref{eq:gen1})-(\ref{eq:gen2})
uses the double series expansion, developed in
\cite{Sneddon,Fabrikant1} and used in \cite{HS} in the context of
a small opening asymptotic. In Appendix A, the general solution is
given. By using these results, we have the following expression
for the coefficient $c_0$
 \beq\label{eq:c0p}
 c_0=-2\alpha\Big[2\log{\frac 1
 \eps}+2\log2+4\beta^2+O(\beta^2,\eps)\Big].
 \eeq
Using equation (\ref{eq:c0}) and (\ref{eq:c0p}), we get the
expression \beq \label{alpha}
\alpha=-\frac1{\Big(\log(\frac1{\beta})+\Big[2\log{\frac 1
\eps}+2\log2+4\beta^2+O(\beta^2,\eps)\Big]\Big)}. \eeq To remember
that $\alpha$ depends on $\beta$ and $\eps$, we denote it
$\alpha(\beta,\eps)$.
For $\epsilon$ fixed and $\delta$ small, the other coefficients are
estimated by using the expression of $c_n$ (given in the appendix)
by \beq c_n=\alpha(\beta,\eps) \tilde{c}_n=O(\alpha(\beta,\eps)),
\eeq where $\tilde{c}_n$ depends only on $n$ and \beq
a_n &= & \frac{c_n}{1+\beta^{2n}} \sim O(\alpha(\beta,\eps))  \\
b_n &=&-c_n\beta^{n} \sim O(\alpha(\beta,\eps)\beta^{n}). \eeq
These estimates show for $\delta$ small that the leading term of
$q$ is given by : \beq\label{eq:asymptotic q} q_{\delta}(r,\theta)
= \left\{
\begin{array}{cc}
1+\alpha(\beta,\eps) \log(r/\delta) +O(\beta) & \hbox{ for } r \sim \delta \\
&  \label{midd}\\
1+\alpha(\beta,\eps) \log(r/\delta)+ O(\alpha) & \hbox{ for } r \sim
R.
\end{array}
\right. \eeq The probability $p_{\delta}(\x)$ that  $T_x^a<T_x^S $
is given by $p_{\delta}(\x)=1-q_{\delta}(\x)$ and the average over a
uniform distribution in formula (\ref{eq:series}) using (\ref{midd})
gives \beq \label{asym}
m_{\delta}=\mean{p_{\delta}}&=&\frac{\int_{A(\delta)} p_{\delta}(r,\theta)rdrd\theta}{\int_{A(\delta)} rdrd\theta}\nonumber \\
&=& -2\frac{\int_{\delta}^R\alpha(\beta,\eps) \log(r/\delta) rdr}{R^2-\delta^2} +O(\beta)\nonumber\\
&=& -\alpha\log{\frac 1 \beta} +O(\beta)\\
&=& \frac{\log{\frac 1 \beta}}{\log{\frac 1 \beta} +2\log{\frac 1
\eps} +2\ln2} +o(1). \nonumber \eeq

In the appendix some properties of $p_{\delta}$ are given when
$\delta$ is small, which will be useful for the next section. From
expression (\ref{asym}), we  obtain the following asymptotic
expression for the mean and the variance of the number of bounds
 \beq \label{mbbp}
 M_b &=&\frac{1-m_{\delta}}{m_{\delta}}= \frac{2\log{\frac 1 \eps} }{\log{\frac 1 \beta}} +O(1)\\
V_b &=&\left(\frac{1-m_{\delta}}{m^2_{\delta}}\right)=2 \left(
\frac{ (\log\frac 1 \eps) (\log{\frac 1 \beta} +2 \log{\frac 1
\eps}+2\log{2})}{\ds(\log{\frac 1 \beta})^2 } \right)+O(1). \eeq
These expressions are valid for $\eps<<1$ fixed, but are uniform in
$\beta$ for $\beta<<1$.
\section{Estimation of Mean First Passage Time \\ $E\{T^A_{\x}|T^A_{\x}<T^S_{\x}\}.$}
In this section, we give an asymptotic estimate of the mean time
to hit $\p \Omega_a$, denoted by $T^A_{\x}$, conditioned on the
event that $ \{ T^A_{\x}<T^{S}_{\x} \}$. As described in
\cite{Karlin}, the conditional process of a Brownian motion
conditioned on $ \{ T^A_{\x}<T^{S}_{\x} \}$ satisfies the
following stochastic differential equation \beq
dX^*(t)=2D\frac{\nabla p_{\delta}(X^*(t))}{
p_{\delta}(X^*(t))}dt+\sqrt{2D}dW. \eeq On the outer boundary $\p
D(R)$, the process $X^*$ is reflected (resp. absorbed) exactly
where $X$ is reflected (resp. absorbed). We denote $t^A(\x)=
E\{T^A_{\x}|T^A_{\x}<T^S_{\x}\}$. It satisfies Dynkin's equation
\cite{Schuss} which here is a degenerated elliptic partial
differential equation, with mixed boundary values:
\beq\label{eqfd}
D p_{\delta} \Delta t^A +2D\nabla t^A\cdot \nabla p_{\delta}&=&-p_{\delta}\mbox{ on} A_{\delta} ,\\
 \frac{\partial t^A }{\partial n}(\x)&=&0\mbox{ on }\partial \Omega _r ,\nonumber \\
 t^A (\x)&=&0\mbox{ on }\partial \Omega _a , \nonumber
\eeq where $\partial \Omega _r $, $\partial \Omega _a $ are
respectively the reflecting part and the absorbing part of the outer
boundary. We remark that no boundary conditions are needed to be
given in the inner circle $\partial D(\delta)$ because
 $\nabla p\cdot n=\frac{\p p}{\p n}>0$ and this is exactly the Fichera conditions,
 discussed in \cite{Fichera,Nirenberg}, where
boundary conditions cannot be given.
\subsection{Asymptotic expansion of the mean first passage time}
When the radius $\delta $ of the inner circle is small, we obtain
an explicit asymptotic expansion of the mean conditional time
$t^A$, solution of equation (\ref{eqfd}). We first derive an
asymptotic solution by gluing two solutions: 1) when the initial
point $\x$ is far from the inner-circle, the solution looks like
the mean exit time when the drift term is set to zero. This
solution is called the outer-solution and has been estimated in
\cite{SSH2} with similar boundary conditions. 2) When the initial
point $\x$ is chosen near the inner-circle, the solution can be
approximated by a radial function. The approximation is valid in a
boundary layer and has to
 match the radial part of the outer-solution at least $C^1$ .
\subsection{Outer-solution}
We now provide a construction of the outer-solution to equation
(\ref{eqfd}). We start with the expansion of $p$, which depends on
the parameter $\delta$.
 \beq p_{\delta}(r,\theta)=
1-\alpha(\delta,\epsilon)\phi_{\delta}(r,\theta). \eeq By using
the appendix, we obtain the following expression \beq
\phi_{\delta}(r,\theta) =\Big( \sum_{n=1}^{\infty}
\frac{\tilde{c_n}}{1+\beta^{2n}}\Big[\Big(\frac r R\Big)^n
-\Big(\frac{\delta \beta} r\Big)^n\Big]\cos(n\theta)+
\log\Big(\frac rR\Big)+4\log(\frac1 \varepsilon)\Big). \eeq
Equation (\ref{eqfd}) can be written outside the boundary layer:
$U_{\delta}=\{ r| r_{\delta} \leq r \leq R\}$, \beq\label{eqfdp1}
& D (1-\alpha(\delta,\epsilon)\phi_{\delta}(r,\theta) ) \Delta t^A
-2\alpha(\delta,\epsilon) \nabla \cdot\phi_{\delta}(r,\theta)
\nabla t^A =-(1-\alpha(\delta,\epsilon)\phi_{\delta}(r,\theta) )
\mbox{ on
} U_{\delta} ,\nonumber\\&\\
& \frac{\p  }{\partial n}t^A(\x)=0 \mbox{ on }\partial \Omega _r ,
\mbox{ and }
 t^A (\x)=0\mbox{ on }\partial \Omega_a. \nonumber
\eeq We look for a regular asymptotic expansion of the solution :
\beq \label{expa} t^A(\x)=u(\x) -\alpha(\delta,\epsilon) u_{1}(\x) +
O(\alpha^2(\delta,\epsilon) ). \eeq Using expression (\ref{expa})
and
 the behavior of $p$ as $\delta$ goes to zero,  in the closed domain $D(R)$ (see appendix), we
obtain from equation (\ref{eqfdp1}) that $u$ satisfies
\beq\label{eqfdp3}
D \Delta u &=&-1\mbox{ on } D(R) ,\\
 \frac{\partial u }{\partial n}&=&0\mbox{ on }\partial \Omega _r ,\nonumber \\
  u&=&0\mbox{ on }\partial \Omega _a.\nonumber \eeq
We can now use the result of \cite{HS,SSH2} to compute the leading
order term of $u$. For $\x$ that does not belong to a boundary
layer near the absorbing boundary $\p\Omega _a$, the asymptotic
expansion of $u$ is given by  \beq u(\x)= \frac{R^2}{D} \big(
\ln(\frac{1}{\varepsilon})+ \ln2 \big)+ O(\varepsilon) \mbox{ on }
U_{\delta}. \eeq
 We conclude that $u$ does not depend on the variable $r$ and $\theta$ at the first order
in $\delta$ and $\epsilon$. Because $u$ is a solution of equation
(\ref{eqfdp3}), outside a boundary layer of $\p\Omega _a$,    the
derivatives $\frac{\p u}{\p r }$ and $\frac{\p u}{r\p \theta }$
are small at the first order in $\delta$ and $\epsilon$. We now
consider the first order term in $\alpha$ in equation
(\ref{eqfdp1}), which satisfies the equation \beqq & D \Delta u_1
-2\nabla \phi\nabla u = 0\\ &\frac{\partial u_1 }{\partial
n}(\x)=0\mbox{ on }\partial \Omega _r, \mbox{ and }
 u_1 (\x)= 0\mbox{ on }\partial \Omega _a.
  \nonumber
\eeqq Outside the boundary layer near $\partial \Omega _a$, the
leading order term  is given by $ u_1 (\x)= o(1)$ (for the variable
$\delta$). We get that \beq \label{ext}
t^A(\x)&=& u(\x) +o(\alpha(\delta,\epsilon) ) \mbox{ on } U_{\delta} \nonumber \\
       &=& \frac{R^2}{D} \big( \ln(\frac{1}{\varepsilon})+ \ln2 \big)
        +o(\alpha(\delta,\epsilon) ) \mbox{ on } U_{\delta}.
\eeq
\subsection{Asymptotic solution inside the boundary layer }
We now provide an asymptotic expansion of the mean time $t^A$
inside the
 boundary layer $BL_{\delta}=\{\delta<r<r_{\delta}\}$, where
$r_{\delta} = \delta(1-\frac1{\alpha(\delta,\epsilon)})$. The size
of the boundary layer is given by formula (\ref{bbll}) in the
appendix. To estimate $t^A$, we use the expression of the
conditional probability $p_{\delta}$. As described in the appendix
(\ref{Boundaryl}), it can be approximated by a radial function, so
that equation (\ref{eqfd}) becomes \beq\label{eqfdp} D \left(
(t^{A})''+\frac{1}{r}(t^A)'\right) +\frac{2D}{p_{\delta}}
{(t^{A})}' \frac{\p p_{\delta}}{ \p r }&= &-1\mbox{ on }
BL_{\delta}, \eeq and the function $\tau$ has to be glued
continuously at $r=r_{\delta}$, using that \beq
p_{\delta}(r,\theta) = -\frac{\alpha(\beta,\eps)}{\delta} \Big(r-
\delta \Big) +\alpha(\beta,\eps)O\Big(r- \delta \Big). \eeq The
leading order term is  \beq \frac{2D}{p_{\delta}}\frac{\p
p_{\delta}}{ \p r }= \frac{2D}{r-\delta}, \eeq
we have \beq\label{eqfd2} D ((t^A)''+\frac{1}{r}(t^A)'
+2\frac{1}{r-\delta}(t^A)') &=&-1\mbox{ on } BL_{\delta}. \eeq The
solution is given for $r> \delta$ by\beq (t^A)'(\x)
=\frac{1}{r(r-\delta)^2}\left(C
-\frac{1}{D}(\frac{r^2\delta^2}{2}-2\delta
\frac{r^3}{3}+\frac{r^4}{4}) \right) \eeq where the constant is
chosen such that the function $t^A$ is integrable. Thus  the
numerator vanishes for $r=\delta$. Actually $r=\delta$ has to be a
third order zero of the numerator and \beq (t^A)'(\x)
=-\frac{r-\delta}{4Dr} \left(r+\delta/3\right). \eeq Thus, \beq
t^A(\x) =-\frac{1}{4D} \left(\frac{r^2}{2}
-\frac{2r\delta}{3}-\frac{\delta^2}{3} \ln(r) +C \right), \eeq
where the constant $C$ is determined by the matching condition:
\beq t^A(r_{\delta}) =\frac{R^2}{D} \big(
\ln(\frac{1}{\varepsilon})+ \ln2 \big)+O(\alpha^2(\delta,\epsilon)
). \eeq We get \beq C \approx
-4R^2\left(\ln(\frac{1}{\varepsilon})+ \ln2\right)
-\frac{\delta^2}{2\alpha^2}, \eeq and for $\x \in BL_{\delta}$,
\beq \label{BL}t^A(\x) &=&\frac{1}{D}
\left(R^2\left(\ln(\frac{1}{\varepsilon})+ \ln2
\right)+\frac{\delta^2}{8\alpha^2}- \frac{r^2}{8}
+\frac{r\delta}{6}+\frac{\delta^2}{12} \ln(r)   \right).\nonumber\\
\eeq
\subsection{Asymptotic estimate of the average Mean time $\mean{\tau}$}
We now compute asymptotically the mean time $\mean{\tau}$ defined in
equation (\ref{eq:tau}) by \beq\label{eq:taur} \mean{\tau}=
\frac{\int_{A_{\delta}} t^A(\x) p_{\delta}(\x)
d\x}{\int_{A_{\delta}}p_{\delta}(\y)d\y} =
\frac{\mean{\tau}_{A_{\delta}}}{m_{\delta}}, \eeq
 where we recall that $
\mean{\tau}_{A_{\delta}} = \frac{1}{Vol({A_{\delta}})}
\int_{A_{\delta}} t^A(\x) p_{\delta}(\x) d\x.$
$Vol({A_{\delta}})=\pi(R^2-\delta^2)$. To compute
$\mean{\tau}_{A_{\delta}}$, we decompose
 the domain $A_{\delta}= BL_{\delta} \cup \left(A_{\delta}-BL_{\delta}\right)$. Using
the previous computations for the outer solution (\ref{ext}) and
the boundary layer solution (\ref{BL}), we compute each term
separately and we get: \beqq \mean{\tau}_{A_{\delta}} =
m_{\delta}\frac{R^2}{D} \big( \ln(\frac{1}{\varepsilon})+ \ln2
\big)+\frac{R^2\beta^4}{9\alpha^2D}+
\frac{R^2}{8D(-\alpha)^3}\ln(\frac{-1}{\alpha})R^2\beta^4+\frac{-\delta^2\alpha}{12DR^2}
\ln(\frac{-\delta}{\alpha})\ln(\frac{-1}{\alpha}).\eeqq
 By taking into account the leading order term only, using the expressions of
$m_{\delta}$ and $p$, we obtain after some computations that \beq
\label{tau}\mean{\tau}\approx\frac{R^2}{D} \big(
\ln(\frac{1}{\varepsilon})+ \ln2 \big) -
\frac{R^2}{8D}\ln(\ln(\frac{1}{\beta}))\ln^3(\frac{1}{\beta})\beta^4
.\eeq We conclude that the boundary layer has very little influence
on the leading order term at the order $\alpha$.
\subsection{Computation of the mean time $\mean{T}$ for a molecule to hit the boundary $\p D(\delta)$ before exit}
We now provide an estimate of mean time it takes for a Brownian
molecule to enter into the binding site domain $D(\delta)$, before
exit. To estimate the mean time $\mean{T}$, we observe that
$E\{T^S_{\x}|T^S_{\x}<T^A_{\x}\}q_{\delta}(\x)+E\{T^A_{\x}|T^A_{\x}<T^S_{\x}\}p_{\delta}(\x)$
is the mean time it takes for a Brownian molecule to exit the domain
$\Omega$ when it can either be absorbed in the inner disk or in the
small absorbing boundary on the outer circle. If we denote the mean
time by \beq\label{meantime}
 w(\x)=E\{T^S_{\x}|T^S_{\x}<T^A_{\x}\}q_{\delta}(\x)+E\{T^A_{\x}|T^A_{\x}<T^S_{\x}\}p_{\delta}(\x)
\eeq then  $w$ satisfies the following PDE \cite{Schuss},
 \beqq
 \Delta w&=&-\frac{1}{D} \mbox{ in } \Omega\\
 w&=&0\quad \mbox{on } \p \Omega_a \cup \p \Omega_\delta \\
 \frac{\p w}{\p n}&=&0\quad \mbox{on }\p \Omega_r\\
 \eeqq
 We substitute $w(r,\theta)=v(r,\theta)+\frac{R^2-r^2}{4D}$,
 and solve the problem for $v$, solution of
 \beqq
 \Delta v=&0&\quad\mbox{ in } \Omega\\
 v=&0&\quad \mbox{on } \p \Omega_a \\
 \frac{\p v}{\p n}=&\frac{R}{2D}&\quad\mbox{on }\p \Omega_r\\
v=-&\frac{R^2-\delta^2}{4D}&\quad\mbox{on }\p \Omega_\delta.
 \eeqq
We proceed as in the previous section by expanding $v$ in Fourier
Series\beq &&v(r,\theta)=\frac{a_0}{2}+\sum_{n=1}^{\infty}
\Big[a_n\Big( \frac r R\Big)^n+b_n\Big(\frac R
r\Big)^n\Big]\cos(n\theta)+\gamma \log\Big(\frac r
R\Big).\label{eq:seriesv}\eeq In a similar way as we estimated $p$
and $q$ (described in appendix A), from the boundary conditions on
$\Omega_\delta$ we get
 \beq
\frac{a_0}{2}+\sum_{n=1}^\infty \left(a_n\beta^n+b_n\left(\frac 1
\beta\right)^n\right)\cos(n\theta)+\gamma \log
\beta=-\frac{R^2-\delta^2}{4D} ,\eeq from which we get \beq
a_0&=&-\frac{R^2-\delta^2}{2D}-2\gamma \log\beta \label{eq:c0 1}\\
-a_n\beta^{2n}&=&b_n.\nonumber\eeq With the definitions $c_0=a_0$,
$c_n=a_n(1-\beta^{2n})$, $H_n=\frac{2\beta^{2n}}{1-\beta^{2n}}$
the boundary conditions on $\p\Omega_r$,$\p\Omega_a$ have the form
of double series equations
\beqq \frac{c_0}{2}+\sum_{n=1}^\infty
\frac{c_n}{1+H_n}\cos(n\theta)=&0,&\quad\theta\in[\pi-\eps,\pi]\\
\sum_{n=1}^\infty n c_n\cos(n\theta)=&
\left(\frac{R^2}{2D}-\gamma\right),&\quad\theta\in[0,\pi-\eps].
\eeqq The solution of these equations follows the steps described in
the appendix and the asymptotic approximation of the coefficient
$c_0$ is given by \beq
c_0=2\left(\frac{R^2}{2D}-\gamma\right)\left(2\log \frac 1 \eps
+2\log2 +O(\beta^2)\right). \eeq Using the last equation and
(\ref{eq:c0 1}) we obtain an equation for $\gamma$
$$-\frac{R^2-\delta^2}{2D}-2\gamma \log\beta=2\left(\frac{R^2}{2D}-\gamma\right)\left(2\log \frac 1 \eps
+2\log2 +O(\beta^2,\eps)\right),$$ which gives that\beq\label{gamma}
\gamma=\frac{R^2}{2D}\frac{4\log\frac 1
\eps+4\log2+1+O(\beta^2,\eps)}{2\log\frac 1 \beta +4\log\frac 1
\eps+4\log2+O(\beta^2)},\eeq and \beq c_0 =
\frac{R^2}{D}\frac{(2\log\frac 1 \beta-1)(2\log\frac 1 \eps
+2\ln2)}{2\log\frac 1 \beta +4\log\frac 1 \eps+4\log2+O(\beta^2)}.
\eeq
\subsection*{Remarks}
Averaging $w$ with respect to a uniform distribution
 \beqq
\mean{w}&=&\frac{1}{\mbox{Vol}(A_\delta)}\int_0^{2\pi}\int_\delta^R w(r,\theta)rdrd\theta\\
&=&\frac{1}{\mbox{Vol}(A_\delta)}2\pi\int_\delta^R
\left(v(r,\theta)+\frac{R^2-r^2}{4D}\right)rdr =
\frac{c_0-\gamma}{2} + \frac{R^2}{8D} \\ &=&\frac{R^2}{2D}\left(
\frac{(2\log\frac 1 \beta-1)(\log\frac 1 \eps+ \log 2)
-\frac12\log\frac 1 \eps - \frac12\log 2+\frac{1}{4}\log\frac 1
\beta -\frac{1}{4}+O(\beta^2,\eps)}{\log\frac 1 \beta +2\log\frac 1
\eps+2\log2+O(\beta^2)} \right).
 \eeqq
 When $\beta\to 0$, we obtain the mean time to exit for only one
  small hole located on the boundary of the domain: this result agrees with
  the result obtained in (\cite{SSH2}, eq.(1.3)).
In the limit $\delta<<\varepsilon$, we have the following
approximation for $\mean{w}$,
 \beq \label{ap1}
 \mean{w}\approx \frac{R^2}{D}\{(\log\frac 1
\eps+ \log 2)+\frac18\},\eeq while for $\varepsilon<<\delta$, we
have \beq \label{ap2}
 \mean{w}\approx \frac{R^2}{2D}\{(\log\frac 1
\beta-\frac12)-\frac14\}. \eeq
\subsection*{Final computations}
To compute $E(\tau_D)$ it is sufficient to estimate the average
$q(\x)E\{T^S_{x}|T^S_{\x}<T^A_{\x}\}$ defined by equation
(\ref{eq:T}) \beq\label{eq-T} \mean{T}= \frac{\int_{A_{\delta}}
E\{T^S_{\x}|T^S_{\x}<T^A_{\x}\}q (\x) d\x}{\int_{A_{\delta}}q
(\y)d\y }. \eeq
 Starting from equation (\ref{meantime}), we get
\beqq q_{\delta}(\x)t^S(\x) &=&w(\x)-p_{\delta}(\x)t^A(\x),
\eeqq thus, \beq \mean{T}=\frac{\int_{A_{\delta}}
t^S(\x)q_{\delta}(\x) d\x}{|A_{\delta}| (1-m_{\delta})}=
\frac{\mean{w}}{1-m_{\delta}}
-\mean{\tau}\frac{m_{\delta}}{1-m_{\delta}}. \eeq Using expression
(\ref{asym}), we have \beq
\frac{1}{|A_{\delta}|}\int_{A_{\delta}}q_{\delta} (\y)d\y
=1-m_{\delta}=\frac{2\ln(\frac{1}{\epsilon})+2\ln
2}{\ln(\frac{1}{\beta})+2\ln(\frac{1}{\epsilon})+2\ln 2}+O(1).
\eeq Finally, the expression of $\mean{T}$ can be computed from
(\ref{ap1}) and (\ref{tau}) and in the limit
$\delta<<\varepsilon$, the leading order term is
\beq \label{T1}
 \mean{T}\approx \frac{R^2}{16D(\log\frac 1 \eps+\ln 2)}\{\log\frac 1
 \beta \} \hbox{ for } \delta<<\varepsilon.
\eeq
To derive the final expression for the Dwell time, we gather
the computational results: using equation (\ref{mbb}), we have
\beq \frac{ 1-m_{\delta}}{m_{\delta}} \approx
\frac{2\ln(\frac1{\epsilon})+2\ln2}{\ln(\frac1{\beta})}.\eeq
Finally,  we obtain from equation  (\ref{eq:formula_mean_mean})
that the dwell time in the approximation $\delta<< \varepsilon$,

\beq \label{dwell-f}
E(\tau_D) &=&\mean{\tau}
+\frac{1-m_{\delta}}{m_{\delta}}(
\mean{T}+\frac{1}{k_{-1}})\nonumber\\ & &  \\
 &\approx& \frac{R^2}{D}
(\ln(\frac{1}{\varepsilon})+\ln
2)+\frac{2\ln(\frac{1}{\epsilon})+2\ln 2}{\ln(\frac{1}{\beta})}
\left( \frac{1}{k_{-1}} + \frac{R^2}{16D(\log\frac 1 \eps+\ln
2)}\{\log\frac 1 \beta \}\right)+o(1).\nonumber \eeq The expression
of $E(\tau_D) $ when $\epsilon<<\delta$ is more delicate and
involves a different boundary layer analysis than the one given in
the appendix.
\section{Discussion and conclusion}
In this article, we have computed the mean time spent by a
molecule inside a microdomain $\Omega$, when it can interact with
some binding sites, represented as a connected sub-domain
$D(\delta) \subset \Omega $. This simplified assumption ignores
the scattered distribution of the binding sites but leaves one
free parameter $\delta$. A rational way to choose the radius
$\delta$ is to equal the length $2\pi \delta$ with the sum of the
potential well boundary length of each binding site.

More generally, the problem of estimating the Dwell time of a
molecule in $\Omega$ when there are already many other independent
molecules, is much more involved. The reason is that even if these
molecules are not interacting, they are coupled through the
competition for the binding sites. However, if the mean number of
bound receptors is fixed and the variance is small enough, the
Dwell time formula (\ref{dwell-f}) can be applied, by choosing for
$\delta$, a value that is related to the amount of free binding
sites available (see below). In general,  when the number of
bounded molecules is fluctuating, a different model is needed to
account for the fluctuations. In that case, a different derivation
of the Dwell time is needed.
\subsection{Receptor trafficking at synapses}
The explicit expression of the Dwell time can be used to describe
receptor dynamics at synapses. A synapse is a micro-contact between
two neurons, involved in signal transmission. In the past decades
experimental observations have revealed that the molecular
composition of the postsynaptic part of a synapse depends on the
history of the neuronal activity
\cite{Nicoll,Bredt,Ehlers1,Choquet,Choquet1, Triller,Lee}.
Although it is a difficult problem to predict the chemical
organization of a synapse, it has been found experimentally that
the type and the number of receptors do not appear randomly, but
are well regulated during specific synaptic plasticity protocols,
such as Long Term Potentiation (LTP). During LTP \cite{Nicoll},
the number and the type of receptors can change, whereas
extra-synaptic receptors diffuse inside a specific microdomain
called the postsynaptic density (PSD). Recently, the concept has
emerged that receptors are constantly moving on the neuronal
surface, in and out of the synaptic domain \cite{Triller}.
Moreover, receptors are  also cycling between the cell surface and
intracellular pools.  According to some recent experimental
results \cite{Choquet}, the movement of the receptors at synapses
can be approximated by a random walk in the heterogeneous PSD.
Moreover, it has been observed that receptors can become confined
for random times \cite{Choquet} and can also be bound to
scaffolding molecules. The PSD delimits a bounded domain $\Omega$
containing small openings at the boundary, where receptors can be
exchanged with the rest of the dendrite. The PSD contains many
fundamental molecules, scaffolding proteins, receptors, kinases,
and many others required for the normal synaptic activity. The PSD
is thus a place where the synaptic molecular machinery is
concentrated. Scaffolding molecules may be used to anchor
receptors and/or to change the biophysical properties of
receptors. Because scaffolding molecules are scattered inside the
domain $\Omega$, in the present model we have replaced the complex
organization of the PSD by a homogenized domain (Figure 1). We can
apply the result of the first part of the present paper to obtain
some estimates of the mean time spent by a receptor inside the
PSD, when it can bind with scaffolding molecules. Estimate
(\ref{dwell-f}) of the Dwell time $E(\tau_D)$ depends on the size
$\delta$ of the scaffolding domain and is a good approximation
when the amount of free scaffolding molecules is small $\delta<<1$
and when the corral barrier, measured by $\varepsilon$, made by
the impenetrable molecules around the PSD, is also small.

\subsection{Dwell time of a single receptor in a synapse containing many other receptors}
To estimate the mean time a receptor stays inside the PSD, we
consider the case where the influx J of receptors entering the PSD
is fixed. The Dwell time can be computed from formula
(\ref{dwell-f}) once the value of the radius $\delta$ is known.
$\delta$ is indeed the radius of the disk containing the free
scaffolding molecules. To obtain an estimate of $\delta$, we
consider the dynamics where each injected receptor can escape with a
rate $\tau$, approximated by formula (\ref{HolcSchuss}). When the
number of injected receptors is balanced by the exiting one from the
PSD, the total number of free receptors  inside the PSD is given by
\beq
[R]=J \tau.
\eeq
(for a detailed analysis see \cite{HT}). Under the steady state
assumption, the number of free scaffolding molecules can be
estimated by using the law of chemical reaction (\ref{chem}) which
says that
\beq\label{chemReact}
K=\frac{[R-S]}{[R][S]},
\eeq
where $[R-S]$ is the number of bounds. $[R]$ (resp. $[S]$) the
number of free receptors (resp. scaffolding molecules) $K$ is the
equilibrium reaction constant per unit area. If $[S_0]$ denotes the
initial number of scaffolding molecules, the conservation of matter
(the total number of scaffolding molecules is conserved) and
equation
\ref{chemReact} implies that
\beq \label{conser}
[S_0]&=&[S]+[R-S]=[S]+K [R][S]\\
      &=&(1+K J\tau)[S]
\eeq
Thus, if $[S]$ (resp. $[S_0]$) occupies a surface $\pi
\delta^2$, (resp.  $\pi \delta_0^2$), then
\beq \label{delta}
\delta = \frac{\delta _0}{\sqrt{1+K J \tau}}.
\eeq
$\delta_0$ can be related to the concentration $c_0$ of scaffolding
molecules by $c_0=\frac{[S_0]}{\pi \delta_0^2}$. To obtain the mean
time estimate, expression (\ref{delta}) of $\delta$ should be used
in the Dwell time formula.

Expression (\ref{delta}) depends on the equilibrium constant $K$,
which in fact depends implicitly of $\delta$. Thus to provide a more
accurate value of $\delta$, we note that the forward binding rate
$k_1$ represents the rate of arrival of a free receptor to a free
binding site and can be approximated by
\beq
 k_1 \approx\frac{1}{\mean{T}},
\eeq
where $\mean{T}$ is given by equation (\ref{eq-T}). Using the
previous considerations and the conservation law \ref{conser}, we
have
\beq \label{conserp}
[S_0]&=&[S]+[R-S]=[S]+K [R][S]\\
      &=&(1+\frac{J\tau}{k_{-1}\mean{T}})[S],
\eeq
where $\tau$ is approximated by expression (\ref{HolcSchuss}) and
$\mean{T}$ (\ref{T1}) depends on $\delta$, which is denoted by
$\mean{T}(\delta)$. Using equation (\ref{delta}), $\delta$ is
solution of equation
\beq \label{eq-delta}
\delta = \frac{\delta _0}{\sqrt{1+\ds{\frac{J \tau}{k_{-1}\mean{T}(\delta)}}}}.
\eeq
In practice, the value of $\delta$ obtained by solving equation
(\ref{eq-delta}) can now be used to estimate the Dwell time, the
mean number of bounds and the other mean times.

Finally, when the total number of receptors inside $\Omega$ is
small, the radius $\delta$ cannot be approximated by a constant,
rather it is fluctuating and in that case, the present approach
needs to be adapted. A variant model of the PSD has been proposed in
\cite{HT}, based on a Markovian approach, which uses an in- and
out-flux of receptors. At steady state, under some approximations,
some estimates of the mean and the fluctuation of the total number
of bound receptors are obtained, however the computations depend
on an a priori expression of the forward binding rate.

\section{Appendix}
\subsection{Solution of double series equation}
In this section we give the mathematical details needed to solve
the double series equation (\ref{eq:gen1})-(\ref{eq:gen2}). The
method was already used in \cite{SSH2},\cite{Sneddon}. Starting
with the equations
 \beq
&\frac{c_0}{2}+\sum_{n=1}^\infty \frac{c_n}{1+H_n}
\cos(n\theta) =0,&\theta\in [\pi,\pi-\eps]\label{eqA:gen1}\\
&&\nonumber\\
 &\alpha+\sum_{n=1}^\infty n c_n
\cos(n\theta)=0,&\theta\in[0,\pi-\eps],\label{eqA:gen2} \eeq we are
interested in computing  $\alpha$  and the coefficients $\{c_i\}$.
Let us define the function $h(\theta)$  for $\theta\in[0,\pi-\eps]$
by
\begin{equation}
\frac{c_0}{2}+\sum_{n=1}^\infty \frac{c_n}{1+H_n}
\cos(n\theta)=\cos
(\frac{\theta}{2})\int_\theta^{\pi-\eps}\frac{h(t)}{\sqrt{\cos(\theta)-\cos(t)}}
\end{equation}
Using this definition and the Fourier coefficient formula
(\cite{SSH2})we can write
\beq &c_n& = \frac{1 + H_n}{\sqrt 2} \int_0^{\pi-\eps} h(t)
[P_n(\cos (t)) + P_{n-1}(\cos(t))] dt,\label{eq:cn}\\
&c_0&=\sqrt{2}\int_0^{\pi-\eps}h(t)dt,\label{eq:c00}\eeq where
 Mehler's identity for Legendre polynomials gives
(\cite {Sneddon} ch.2)
\[P_n(\cos(u))=\frac{\sqrt {2}}{\pi}\int_0^u\frac{\cos(n+\frac{1}{2})d\theta}{\sqrt{\cos(\theta)-\cos(u)}}.\]
Integrating equation (\ref{eqA:gen2}), using the expression for
the coefficients (\ref{eq:cn}) into the integrated equation and
the identity \beqq
&\frac{1}{\sqrt{2}}\sum_{n=1}^\infty\{P_n(\cos(t))+P_{n-1}(\cos(t))\}\sin(n\theta)=
\frac{\cos(\frac{1}{2}\theta)H(\theta-t)}{\sqrt{\cos(t)-\cos(\theta)}},
\eeqq where $H$ is the Heaviside function, we get the integral
equation \beq &&\int _0^\theta
\frac{h(t)}{\sqrt{\cos(t)-\cos(\theta)}}+\int_0^{\pi-\eps}K_\beta(\theta,t)h(t)dt=\frac{-\alpha\theta}{\cos(\frac{\theta}{2})},
\label{eqA:int1}\eeq where the kernel function $K_\beta$ is
\[K_\beta(\theta,s)=\frac{1}{\sqrt{2}\cos(\frac{\theta}{2})}\sum_{n=1}^\infty H_n(P_n(\cos(t)+P_n(\cos(t))).\]
Using the first elements in the infinite series we can give an
asymptotic expansion to $K_\beta$
\begin{equation}
K_\beta(t,s)=2\beta^2\cos^2(s)\sin(t)+\mbox{O}(\beta^4).
\end{equation}
Equation (\ref{eqA:int1}) can be transformed according to the Abel
transform  \beqq
f(x)&=&\int_0^x\frac{g(t)dt}{(p(x)-p(t))^q}\\
g(t)&=& -\frac{\sin(\pi
q)}{\pi}\frac{d}{dt}\int_0^t\frac{p'(x)f(x)dx}{(p(x)-p(t))^{1-q}},\eeqq
 where $p(x)$ is monotonic increasing function. Transforming eq.(\ref{eqA:int1}), with $q=\frac{1}{2}$
 and $p(x)=-\cos(x)$ we get
\beq
h(\theta)-\int_0^{\pi-\eps}\tilde{K}_\beta(\theta,t)h(t)dt=\frac{1}{\pi}\frac{d}{d\theta}\int_0^\theta
\frac{2\alpha t \sin(\frac{1}{2}t)dt}{\sqrt{\cos t -\cos
\theta)}}\label{eqA:int2}
 \eeq
with
$\tilde{K}_\beta(t,s)=-\frac{1}{\pi}\frac{d}{dt}\int_0^t\frac{K_\beta(u,s)\sin(u))}{\sqrt{\cos(u)-\cos(t)
}}$. If we define
$z(\theta)=\frac{1}{\pi}\frac{d}{d\theta}\int_0^\theta \frac{2\alpha
t \sin(\frac{1}{2}t)dt}{\sqrt{\cos t-\cos \theta}}$, then equation
(\ref{eqA:int1}) has the form $(I-\tilde{K}_\beta)h=z$. This is the
Fredholm integral equation, which can be approximated using the
infinite sum
\beq
\label{expand}
h=z+\tilde{K}_\beta z +\tilde{K}^2_\beta z+\ldots
\eeq
Computing the first elements in the sum gives an approximation to
$h(\theta)$ which in turn, by substituting the approximation into
the expressions for $c_n$ gives an approximation for the
coefficients. To calculate $c_0$ up to $\mbox{O}(\eps,\beta^4)$ we
have to evaluate the integral
\[\sqrt{2}\left(\int_0^{\pi-\eps}z(t)dt+\int_0^{\pi-\eps}\int_0^{\pi-\eps}\tilde{K(s,t)}_\beta
z(t) dtds\right). \] The computation is made in \cite{SSH2} and
gives
\[c_0\approx 2\alpha (2\log \frac{1}{\eps}+2\log 2+4\beta^2).\]
\subsection{Properties of the conditional probability $p_{\delta}$}
\subsubsection{Inside $A_{\delta}$}
$q_{\delta}$ is solution of equation (\ref{eq5}) and depends on
the parameter $\delta$.
\begin{prop}
The sequence $p_{\delta}$ converges uniformly to the constant 1,
on any compact set $K$ strictly contained in the domain
$A_{\delta}$. More specifically, for any $K \subset A_{\delta}$,
there exists a constant $C>0$ such that for all $r,\theta \in K$,
 \beqq
|p_{\delta}(r,\theta)-1| &\leq& C \alpha(\beta,\eps)  \\ \\
|\nabla p_{\delta}(r,\theta)| &\leq&  C \alpha(\beta,\eps)
 \eeqq
where $\alpha(\beta,\eps)$ is  defined by (\ref{alpha}) and \beq
0<\alpha(\beta,\eps)=\alpha(\frac{\delta}{R},\eps)\leq
\frac{C}{\ln(\frac1{\delta})} \eeq which tends to zero when
$\delta$ goes to zero.
\end{prop}
{\noindent \bf Proof.}
To estimate $p_{\delta}$ we consider expression (\ref{eq:series}),
which can be written as \beqq
&&-p_{\delta}=q_{\delta}(r,\theta)-1=\sum_{n=1}^{\infty}
\frac{c_n}{1+\beta^{2n}}\Big[\Big( \frac r
R\Big)^n-\Big(\frac{\delta \beta}
r\Big)^n\Big]\cos(n\theta)+\alpha(\beta,\eps) \log\Big(\frac r
\delta\Big).\label{eq:seriesp} \eeqq where $\alpha(\beta,\eps)$ is
given by expression  (\ref{alpha}) and we recall that \beq
\label{coef2} &c_n& = \frac{1 + H_n}{\sqrt 2} \int_0^{\pi-\eps}
h(t) P_n(\cos (t)) + P_{n-1}(\cos(t))] dt. \eeq By definition
$H_n=O(\beta^{2n})$ and $P_n$ is the Legendre polynomial. For
$x\in [-1,1]$ and all n, $|P_{n}(x)| \leq 1$. Moreover,  using
formula (\ref{expand}), we obtain that \beq \label{fomzp}
h(t)=z(t)+O(\beta^2)=-\frac{2\alpha}{\pi}\frac{d}{dt}\int_0^t\frac{u
\sin u}{\sqrt{\cos u-\cos t}}du. \eeq Thus we conclude from the
explicit formula (\ref{coef2}) that there exists a constant $C>0$
such that for all $n$ \beq |c_n| \leq C \alpha(\beta,\eps) \eeq
which is independent of $n$. We denote
$c_n=\tilde{c_n}\alpha(\beta,\eps).$ We can now obtain the desired
estimates. For $r_0<r<R_1<R$, we get, \beq \label{express}
1-p_{\delta}(r,\theta)=1+\alpha(\beta,\eps) \Big(
\sum_{n=1}^{\infty}
\frac{\tilde{c_n}}{1+\beta^{2n}}\Big[\Big(\frac r
R\Big)^n-\Big(\frac{\delta \beta} r\Big)^n\Big]\cos(n\theta)+
\log\Big(\frac r \delta\Big)\Big). \eeq Thus, \beq
|1-p_{\delta}(r,\theta)|\leq  I+ II, \eeq where \beq
I \leq |1-\alpha(\beta,\eps)\log \Big(\frac{R_1}\delta \Big)| \leq C \frac{\ln(R_1/r_0)+O(1)}{\ln(R/\delta)+O(1)}\\
II\leq \alpha(\beta,\eps) \sum_{n=1}^{\infty}  \Big[\Big(
\frac{R_1} R\Big)^n+\Big(\frac{\delta \beta} {r_0}
\Big)^n\Big]\leq C \alpha(\beta,\eps). \eeq The last part of these
inequalities uses the asymptotic expression of $\alpha$ given by
formula (\ref{alpha}). These inequalities show that on any compact
set of $A_{\delta}$, $1-p_{\delta}$ converges uniformly to zero at
a rate $\frac1{\ln(1/\delta)}$. Similarly, to estimate the
gradient we estimate separately $|\frac{\p p}{\p r}|$ and
$|\frac1{r}\frac{\p p}{\p \theta}|$. First, \beqq -\frac{\p
p_{\delta}}{\p r}(r,\theta)=\alpha(\beta,\eps) \Big(
\sum_{n=1}^{\infty} \frac{n\tilde{c_n}}{1+\beta^{2n}}\Big[\Big(
\frac{r^{n-1}}{R^n}\Big)+\Big(\frac{(\delta
\beta)^n}{r^{n+1}}\Big) \Big]\cos(n\theta)+ \frac{1}{r} \Big),
\eeqq we get \beqq |\frac{\p p_{\delta}}{\p r}(r,\theta)| \leq
|\alpha(\beta,\eps)| \Big( \sum_{n=1}^{\infty} n \Big[\Big(
\frac{R_1^{n-1}}{R^n}\Big)+\Big(\frac{(\delta
\beta)^n}{r_o^{n+1}}\Big) \Big]+ \frac{1}{r_0} \Big)\leq
C|\alpha(\beta,\eps)| \eeqq and second, \beqq \frac{1}{r}\frac{\p
p_{\delta}}{\p \theta}(r,\theta)=\frac{\alpha(\beta,\eps)}{r}
\Big( \sum_{n=1}^{\infty}
\frac{\tilde{c_n}}{1+\beta^{2n}}\Big[\Big( \frac r
R\Big)^n-\Big(\frac{\delta \beta} r\Big)^n\Big]n\sin(n\theta)\Big)
\eeqq and \beqq |\frac1{r}\frac{\p p}{\p \theta}| \leq
|\frac{\alpha(\beta,\eps)}{r_0}| \Big( \sum_{n=1}^{\infty}
\Big[\Big( \frac {R_1} R\Big)^n+\Big(\frac{\delta \beta}{r_0}
\Big)^n\Big]n\Big)\leq C|\alpha(\beta,\eps)|. \eeqq \QED
\subsubsection{The boundary layer}\label{Boundaryl}
We consider now the behavior of $p_{\delta}$ near the the inner
circle $r=\delta$, where $p_{\delta}$ vanishes. We expect to see a
boundary layer near the circle $r=\delta$, as $\delta$ goes to
zero. $p_{\delta}$  is a regular function and a Taylor expansion
in the $r$-variable inside expression (\ref{express}) gives that
\beqq -p_{\delta}(r,\theta) = \frac{\alpha(\beta,\eps)}{\delta}
\Big(r- \delta \Big) + \alpha(\beta,\eps)  \sum_{n=1}^{\infty}
n\frac{\tilde{c_n}}{1+\beta^{2n}}\Big[ 2R
\beta^{n-1}\Big]\cos(n\theta)\Big(r- \delta \Big)+O\Big(r- \delta
\Big)^2. \eeqq The leading order term is radial and we get
\beq\label{bbl} p_{\delta}(r,\theta) =
-\frac{\alpha(\beta,\eps)}{\delta} \Big(r- \delta \Big)
+\alpha(\beta,\eps)O\Big(r- \delta \Big). \eeq To estimate the
size of the boundary layer, we look at the value $r_{\delta}$ such
that $p_{\delta}$ is of the order one. Using the previous
approximation of $p_{\delta}$ (\ref{bbl}) we get \beq\label{bbll}
r_{\delta} &=&
\delta-\frac{\delta}{\alpha(\beta,\eps)}\\
 &=&  \delta+ \delta(\ln\frac{1}{\delta}+2 \ln\frac{1}{\eps}+2\ln2+\ln
 R+O(\beta^2)).
 \eeq
\subsubsection{Convergence of $p_{\delta}$ in the neighborhood of $\p D(R)$}
The sequence of harmonic function  $p_{\delta}$ converges
uniformly on any compact set of $D(R)-\{0\}$ to the function 1. We
show now that this is also the case in the neighborhood of $\p
D(R)$. We use the elliptic estimates given in \cite{Gilbarg}
derived at the boundary. We can also directly extend as a harmonic
function $p_{\delta}$ to a neighborhood $T$ of $D(R)$. We have
that \beq \sup_{\{x \in K\}} | \nabla p_{\delta}| \leq C \eeq for
any compact $K$ in $T$. By Ascoli's theorem, there exists a
subsequence of $p_{\delta}$ converging in $C^{0,\alpha}$. We
conclude since $p_{\delta}$ converges to 1 on any compact set of
$D(R)-\{0\}$ that the sequence $p_{\delta}$ converges to 1 on
$T-\{0\}$ and is a weak solution of \beq\label{eeqq}
\Delta u&=&0 \mbox{ on } D(R)-{0} \\
    u(x)&=&1 \mbox{ on }\p \Omega_a ,\nonumber\\
\frac{\p u}{\p n}(x)&=& \mbox{ on }\p D(\delta), \nonumber \eeq
Because $0\leq u\leq 1$, the solution can be extended up to 0 and
the only solution of equation (\ref{eeqq}) is the constant 1. We
conclude that the sequence $p_{\delta}$ converges to the constant
one uniformly on $\overline{D(R)}$.\newpage

\bibliographystyle{amsplain}

\end{document}